\documentclass[pra,aps,twocolumn,superscriptaddress,showpacs,showkeys,preprintnumbers,amsmath,amssymb]{revtex4}
\usepackage{graphicx,amsfonts,amsmath}% Include figure files
\usepackage{dcolumn}% Align table columns on decimal point
\usepackage{bm}% bold math

\newcommand{\bk}{b_{\bm{k}}}
\newcommand{\bkd}{b_{\bm{k}}^{\dag}}
\newcommand{\bmk}{b_{-\bm{k}}}

\newcommand{\sumk}{\sum_{\bm{k}}}
\newcommand{\wk}{\omega_{\bm{k}}}
\newcommand{\fk}{f_{\bm{k}}}

\newcommand{\pe}{|1\rangle\!\langle 1|}

\newcommand{\unit}{\mathbb{I}}

\begin{document}

%\preprint{}

\title{Sudden death of effective entanglement}% Force line breaks with \\

\author{K. Roszak}

\affiliation{Department of Condensed Matter Physics,
Faculty of Mathematics and Physics, Charles University,
12116 Prague, Czech Republic}%Lines break automatically or can be forced with \\

\affiliation{Institute of Physics, Wroc\l aw University of Technology, 50-370 
Wroc\l aw, Poland}

\author{P. Horodecki}

\affiliation{Faculty of Applied Physics and Mathematics, Gda\'{n}sk
University of Technology, 80-952 Gda\'{n}sk, Poland}

\affiliation{National Quantum Information Centre of Gda\'{n}sk,
81-824 Sopot, Poland}

\author{R. Horodecki}

\affiliation{National Quantum Information Centre of Gda\'{n}sk, 81-824 Sopot, Poland}

\affiliation{Institute of Theoretical Physics and Astrophysics, University of Gda\'{n}sk, 80-952 Gda\'{n}sk, Poland}

\begin{abstract}

Sudden death of entanglement is a well-known effect resulting from the finite volume 
of separable states. We study the case when the observer has a limited measurement 
capability and analyse the effective entanglement (i.e., entanglement minimized over the
output data). 
We show that in the well defined system of two quantum dots monitored 
by single electron transistors, one may observe a sudden death of effective entanglement when real, physical 
entanglement is still alive. For certain measurement setups,
this occurs even for initial states for which sudden death of physical entanglement
is not possible at all.
The principles of the analysis may be applied to other analogous scenarios, such as
estimation of the parameters arising from quantum process tomography. 

 \end{abstract}

\maketitle
\section{Introduction} 
Sudden death of quantum entanglement \cite{zyczkowski01,rajagopal,yu04,eberly07} is one of the phenomena related 
to the fact that in finite dimensional systems the set of nonentangled states is of finite volume \cite{zyczkowski98}.
The phenomenon was explicitly identified in Ref. \cite{zyczkowski01} (see also its implicit presence 
in independent analysis \cite{rajagopal}). Full recognition of its importance and 
consequences was established with time \cite{yu04,eberly07,ficek06,sainz08} (for the review see Ref. \cite{yu09})
and demonstrated experimentally \cite{almeida07}.

In earlier analysis of entanglement evolution, it was assumed that the observer has the power to perform 
arbitrary measurements and can determine the state of the system completely. 
This is however not always true. In particular, there are natural systems, like quantum dots (QDs)
which we will consider later, where limited measurement power is a natural 
and practical constraint (ie. single electron transistors (SETs) coupled to QD systems 
can be used to find a limited amount of information about the QD state \cite{stace04b,stace04a})
In all such cases of limited measurement capability, it is natural to consider the worst case scenario 
proposed in Ref. \cite{horodecki99}: {\it as real entanglement, one should consider the entanglement 
(i.e. chosen entanglement measure) minimized over the set of measurement data.}
This approach has found wide developments in terms of entanglement 
witnesses \cite{eisert,guhne07a,audenaert06,guhne08} with respect to experimental data
\cite{puentes09,schmid08}.
The minimized entanglement will subsequently be called {\it effective entanglement}. 

Here, we consider the system of a double QD (DQD) interacting with a phonon-bath
which leads to an unavoidable partial
pure dephasing effect typically on picosecond timescales \cite{borri01,vagov03,vagov04}.
The time-evolution 
of physical entanglement in this system can be found in Ref. \cite{roszak06a}.
We take into account the limitations to the knowledge of the system state imposed by a realistic 
measurement setup (consisting of different configurations of SETs 
interacting with the double-dot system). This measurement scheme does not allow
for state tomography and provides, in fact, a very limited set of observables.
In each time-step we minimize the value of entanglement with respect to the data 
which can be measured
and find the evolution of the {\it effective entanglement with respect to the SET-defined observables}.
We show that such an evolution, besides leading to a quantitative reduction of entanglement
(compared to physical entanglement), demonstrates qualitative changes such as the possibility of sudden
death of effective entanglement in situations when physical entanglement lives for
arbitrarily long times. 
Our approach may be extended to estimate the time evolution of quantities
describing quantum processes (e. g., entangling power or 
fidelity of a quantum process).

\section{The model}
The system under study consists of two parts, the measured and the measuring 
subsystems. The former consists of 
the DQD ensemble, in which superpositions of excitonic states undergo
pure dephasing due to the interaction with the phonon modes of the surrounding
crystal.
The pure dephasing is only partial, as has been experimentally shown \cite{borri01}
and later explained theoretically \cite{vagov03,vagov04}; experimental and theoretical data
yield both qualitative and 
quantitative agreement \cite{vagov04}.
The $|0\rangle$ and $|1\rangle$ states of the qubit correspond
to an empty QD and a QD with an exciton in its ground state, respectively
(the two qubits are located in separate QDs).
A number of SETs which determine
the charge distribution in their vicinity constitute the latter.
This allows for the measurement of a set of elements of the DQD
density matrix (the precise elements measured are determined
by the geometry of the measurement device with respect
to the DQD), and the subsequent calculation of effective entanglement.

The Hamiltonian which governs the evolution of the
excitonic states in the DQD is
\begin{eqnarray}
H  &=&  \epsilon_{1}(\pe\otimes\unit)
+\epsilon_{2}(\unit\otimes\pe) \\
\nonumber
&&+(\pe\otimes\unit)\sumk\fk^{(1)}(\bkd+\bmk)\\
\nonumber
&&+(\unit\otimes\pe)\sumk\fk^{(2)}(\bkd+\bmk)
+\sumk\wk\bkd\bk,
\end{eqnarray}
where the two states of each QD are denoted by
$|0\rangle$ and $|1\rangle$, $\mathbb{I}$ is the unit operator,
$\epsilon_{1,2}$ are the transition energies
in the two QDs, $\fk^{(1,2)}$
are exciton-phonon coupling constants, $\bkd, \bk$ are 
creation and annihilation
operators of the phonon modes, and $\wk$ are the corresponding energies
(we put $\hbar=1$). The explicit tensor notation refers to the DQD 
but is suppressed for the phonon reservoir components.

Exciton wave functions are modelled by anisotropic Gaussians
with the extension $l_{\mathrm{e/h}}$ in the $xy$ plane
for the electron/hole, and $l_{z}$
along $z$ for both particles.
Then, the coupling constants for the deformation potential coupling between
confined charges and longitudinal phonon modes have the form
$\fk^{(1,2)}=\fk e^{\pm ik_{z}d/2}$, where
\begin{displaymath}
\fk=\sqrt{\frac{k}{2\varrho Vc}}e^{-l_{z}^{2}k_{z}^{2}/4}
\left[\sigma_{\mathrm{e}}e^{-l_{\mathrm{e}}^{2}k_{\bot}^{2}/4}
-\sigma_{\mathrm{h}} e^{-l_{\mathrm{h}}^{2}k_{\bot}^{2}/4} \right],
\end{displaymath}
$V$ is the normalisation volume of the bosonic reservoir, $d$ is
the distance between the subsystems,
$k_{\bot,z}$ are momentum
components in the $xy$ plane and along the $z$ axis,
$\sigma_{\mathrm{e,h}}$ are deformation potential constants for
electrons and holes, $c$ is the speed of longitudinal sound,
and $\varrho$ is the crystal density.

In our calculations we use parameters typical for two 
self-assembled GaAs/InGaAs QDs stacked on top of each other \cite{roszak06a,roszak09}.
The material parameters used are $\sigma_{\mathrm{e}}=8$ eV,
$\sigma_{\mathrm{h}}=-1$ eV, $c=5.1$ nm/ps, $\varrho=5360$ kg/m$^{3}$
(corresponding to GaAs), and  $d=6$ nm,
$l_{\mathrm{e}}=4.4$ nm, $l_{\mathrm{h}}=3.6$ nm, $l_{z}=1$ nm
(dot related parameters). 

The Hamiltonian can be diagonalised exactly using the 
Weyl operator method \cite{roszak05a} and the evolution is calculated
following Ref. \cite{roszak06a}. The interaction with the phonon modes leads
to partial pure dephasing \cite{borri01,krummheuer02,vagov04}, 
leaving the state occupations unchanged;
the explicit forms of the time dependence of the off-diagonal
density matrix elements may be found in Ref. \cite{roszak06a}.

The measuring subsystem is taken into account only in principle,
in the sense that the information which can be gained
about the state of the DQD ensemble is limited by realistic
measurement capability.
When measuring the state of a DQD by observing
the current through a SET, the actual observable depends
on its position with respect to the DQD. The number
of the DQD density matrix elements which can be obtained
is very limited; this restriction is a key point in this
paper. There are a number of features inherent to this 
measurement scheme, e.g. finite measurement time and 
an uncertainty of the outcome, which are not taken into
account for the sake of clarity.

Let us consider a SET located near the lower QD 
(configuration A).
The interaction between the SET electron and the exciton in the dot
shifts the energy levels in the SET when the QD is occupied, hence
affecting the current. An appropriate choice of SET parameters
allows for the maximisation of the difference in current flow
and the measurement of the occupation of 
the lower dot \cite{stace04b}. 
The measurement projectors corresponding 
to this situation are 
$P_n=|0\rangle\langle 0|\otimes\unit$ and
$P_e=|1\rangle\langle 1|\otimes\unit$,
so the measured quantity is 
\begin{equation}
\label{x}
x=\langle 00|\rho|00\rangle+\langle 01|\rho|01\rangle.
\end{equation}
If such a SET is located near the upper QD,
the measured quantity is 
\begin{equation}
\label{y}
y=\langle 00|\rho|00\rangle+\langle 10|\rho|10\rangle.
\end{equation}

Configuration B involves a SET located symmetrically
between the QDs and in close proximity to them in such a way
that the energy level on the SET island is sensitive to the
probability of finding an exciton in midpoint \cite{stace04a}.
The corresponding projectors are
$P_n=|-\rangle\langle -|$ and
$P_e=|+\rangle\langle +|$, where
\begin{equation}
\label{pm}
|\pm\rangle=(|01\rangle\pm|10\rangle)/\sqrt{2}.
\end{equation}
This allows for the measurement of a linear combination
of density matrix elements
\begin{equation}
\label{z}
z=\langle 01|\rho|01\rangle +\langle 10|\rho|10\rangle
+2\mathrm{Re}\langle 01|\rho|10\rangle.
\end{equation}

If the SET is located further away from the DQD region (symmetrically),
it is sensitive to the total number of excitons (configuration C). 
This allows for the 
measurement of 
\begin{equation}
\label{d}
d=\langle 11|\rho|11\rangle
\end{equation}
via projectors
$P_e=|11\rangle\langle 11|$ and
$P_n=|+\rangle\langle +|+|-\rangle\langle -|+|00\rangle\langle 00|$,
or $a=\langle 00|\rho|00\rangle$ via
$P_e=|00\rangle\langle 00|$ and
$P_n=|+\rangle\langle +|+|-\rangle\langle -|+|11\rangle\langle 11|$
depending on the SET parameters. Switching between the two modes
can be accomplished by applying different voltage to the SET.

\section{Minimising entanglement from the SET data}
Consider the general quantum state $\rho$. 
There is a simple

{\it Lemma .- For any convex entanglement measure $E$ (or entanglement parameter) which 
is invariant under complex conjugate one has
$E(\rho)\geq E(\mathrm{Re}(\rho))$, where the "$\mathrm{Re}$" symbol means 
real part of the quantum state.}

{\it Proof.-} $E(\mathrm{Re}(\rho))=E(\frac{1}{2} \rho + \frac{1}{2} \rho^{*})\leq \frac{1}{2} E(\rho) 
+ \frac{1}{2} E(\rho^{*})=
 E(\rho)$.
 
Any reasonable entanglement measure should be invariant under the complex conjugate;
all known entanglement measures like entanglement of formation, concurrence, 
lognegativity, relative entropy of entanglement 
and all distillable quantities fulfil this condition. 
Some of them are also convex i.e. entanglement of formation, concurrence and 
relative entropy of entanglement.
In what follows we shall consider the concurrence \cite{hill97,wootters98}, 
which is both 
convex and invariant under the complex conjugate. The reason for this choice of entanglement
measure is its mathematical simplicity and the fact that it is easily converted
into Entanglement of Formation (which has a good physical interpretation).  Since the 
SET measurement provides only real parts of off-diagonal density matrix elements,
the first step of minimisation  
is to consider only the real parts of DQD states,
$\tilde{\rho}=\mathrm{Re}(\rho)$. 

\section{Sudden death of effective entanglement}
Firstly, we will consider the situation, when a number of SETs provide 
all possible information that can be gained about the state of the DQD
with this measurement technique.
This requires a pair of SETs in configuration A located near the two QDs,
giving $x$ and $y$ defined, respectively, by Eq. (\ref{x}) and (\ref{y}),
and one in configuration C providing $d$ (Eq. (\ref{d})). Hence, all of the
diagonal elements of the density matrix can be found,
\begin{eqnarray}
\label{abcd}
\langle 00|\rho|00\rangle&=&x+y+d-1\equiv a,\\
\nonumber
\langle 01|\rho|01\rangle&=&1-y-d\equiv b,\\
\nonumber
\langle 10|\rho|10\rangle&=&1-x-d\equiv c,\\
\nonumber
\langle 11|\rho|11\rangle&=&d.
\end{eqnarray}
A SET in configuration B provides $z$ (Eq. (\ref{z})) giving
the real part of one of the off-diagonal density matrix elements
\begin{equation}
\label{reh}
\mathrm{Re}(\langle 01|\rho|10\rangle)=\frac{x+y+z}{2}+d-1
\equiv \mathrm{Re} (h).
\end{equation}

\begin{figure}[tb]
\begin{center}
\unitlength 1mm
\begin{picture}(85,45)(0,0)
\put(0,-8){\resizebox{80mm}{!}{\includegraphics{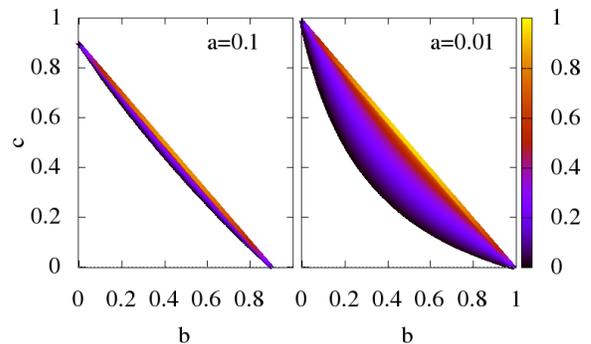}}}
\end{picture}
\end{center}
\caption{\label{fig1} The effective entanglement (colorscale)
of coherent states in the first measurement setup
as a function of known occupations $b$ and $c$
for two values of $a$. The plot shows only nonzero effective
entanglement.}
\end{figure}

Since all of the diagonal density matrix elements are known,
the set of initial maximally entangled states which cannot
exhibit sudden death of effective entanglement is the same
as the set of states with real off-diagonal density matrix elements
which do not exhibit 
sudden death of physical entanglement \cite{roszak06a}. The time evolution
of effective and physical entanglement in these states
under pure dephasing is the same
and the concurrence is equal to 
$C(\tilde{\rho})=2\mathrm{Re}(h)$.
The situation is different for states where all diagonal
density matrix elements are non-zero where the set of effectively entangled
coherent states is substantially reduced. 
In Fig. \ref{fig1} the effective entanglement of coherent states
($\mathrm{Re} (h)=\sqrt{bc}$)
as a function of $b$ and $c$
is plotted for two values of $a$.

\begin{figure}[tb]
\begin{center}
\unitlength 1mm
\begin{picture}(85,45)(0,0)
\put(8,-10){\resizebox{80mm}{!}{\includegraphics{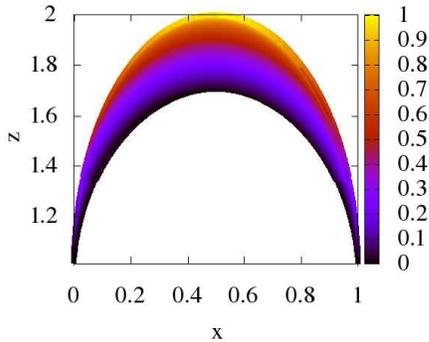}}}
\end{picture}
\end{center}
\caption{\label{fig2} The effective entanglement (colorscale)
in the second measurement setup
as a function of the observables $x$ and $z$.
The plot shows only nonzero effective
entanglement.}
\end{figure}

Secondly, let us consider the situation when only $x$ and $z$ have
been measured (configuration A and B), so only linear combinations
of some density matrix elements are known.
The measurement outcome $x\in [0,1]$. The outcome $z\in[0,2]$,
but it is easy to show that for non-zero effective entanglement $z>1$.
The minimisation of 
entanglement requires that the quantity $ad$ be maximized, which
occurs for $\mathrm{Re} (h)=\sqrt{bc}$ (the unknown density matrix
elements $a$, $b$, $c$, $d=1-a-b-c$ and $\mathrm{Re} (h)$ are
defined in Eq. (\ref{abcd}) and (\ref{reh})); then
$ad=
(x-b) (1 - x - (-\sqrt{b}
+ \sqrt{z})^2).$
Finding the maximal $ad$, which leads to set values for
$a$, $b$, $c$, $d=1-a-b-c$ and $\mathrm{Re} (h)$, and minimising
over the five still unknown off-diagonal density matrix elements
produces the effective entanglement for given measurement outcomes $x$
and $z$. Effective entanglement as a function of $x$
and $z$ is plotted in Fig. \ref{fig2}.

It is interesting to consider here the time-evolution of 
effective entanglement under phonon-induced pure dephasing of an initially
maximally entangled
state $|+\rangle$ (Eq. (\ref{pm})). When measured it
will yield $x=0.5$ and $z=2$, so the effective concurrence 
$C_e(|+\rangle\langle +|)=1$
(note that the state $|-\rangle$ has
$x=0.5$, but $z=0$ and $C_e(|-\rangle\langle -|)=0$).
Phonon-induced evolution of the state does not change
$x$, but $z$ decreases with decreasing $\mathrm{Re} (h)$ leading
to sudden death of effective entanglement for sufficiently dephased
states. This state does
not exhibit sudden death of physical entanglement. 

\begin{figure}[tb]
\begin{center}
\unitlength 1mm
\begin{picture}(85,43)(0,0)
\put(4,-8){\resizebox{70mm}{!}{\includegraphics{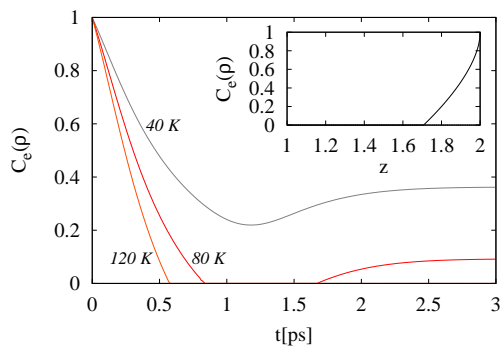}}}
\end{picture}
\end{center}
\caption{\label{fig3} The time-evolution of effective entanglement for different temperatures.
Inset: Effective entanglement as a function of the measured parameter $z$}
\end{figure}

Thirdly, we will consider the simplest situation, where
the measurement device is limited to configuration B
and the measurement data yield only $z$ as defined in Eq. (\ref{z}).
The amount of information gained by the measurement is very limited.
Minimising effective entanglement requires the maximisation
of $ad$ the same as in the previous measurement setup,
yet now the maximum is easily found for $b=c=\mathrm{Re} (h)=z/4$.
The dependence of effective entanglement on $z$ is plotted
in the inset of Fig. \ref{fig3}.

The evolution of effective entanglement
of the initial state $|+\rangle$ in this setup under realistic
phonon-induced pure dephasing is plotted in Fig. \ref{fig3} for
different temperatures. As is to be expected, effective disentanglement
occurs faster than physical disentanglement. Furthermore, sudden death
of entanglement appears for sufficiently high temperatures
(e. g., when the dephasing is strong enough). For a limited
range of temperatures, sudden birth of entanglement is also observed.
The second phenomenon is due to the enhancement of coherence 
which occurs when wavepackets from the two QDs meet 
due to positive interference between them; this mechanism does
not lead to the sudden birth of physical entanglement \cite{roszak06a}.

\section{Conclusion and discussion}

   We have considered the evolution of entanglement in a DQD system under phonon-induced 
pure dephasing minimized over a set of attainable data (the considered measurement schemes
are based on SETs). 
In all of the considered measurement setups a reduction of effective (minimized)
entanglement compared to the physical one was observed. Furthermore,
in the setups where the knowledge of QD occupations is limited,
sudden death was found in situations when
the physical entanglement lives for
arbitrarily long times and its sudden death is not possible. For a limited
range of temperatures (coupling strengths)
even the sudden birth of entanglement occurs (due to a mechanism
that does not cause sudden birth of physical entanglement).
Hence, we have shown that the analysis of entanglement dynamics 
in systems with difficult measurement access can lead to qualitative
differences in measured entanglement. 

  The present analysis may provide a starting point to analogous research in 
general dynamics including quantum process tomography (\cite{chuang97,riebe06}). 
For instance, one may want to probe the entangling power \cite{zanardi00} of a sequence 
of quantum gates in time in a possibly cheap way, when only some of measurements 
are not costly. Then the principles of the above analysis may be applied with the
help of the Choi-Jamio\l kowski isomorphism \cite{jamiolkowski72,choi82}. 
Fundamental parameters of the dynamics,
including the fidelity of a quantum process, can be then estimated, 
i.e. by minimisation in time under a restricted set of data, 
especially when coarse-graining 
information about the dynamics is needed.

\begin{acknowledgments}
This work was supported by the UE IP SCALA project and by the Czech Science
Foundation under Grant No. 202/07/J051. R. H. and P. H. acknowledge support from the
Polish Ministry of Science under Grant No. NN202231937.
\end{acknowledgments}


\begin{thebibliography}{32}
\expandafter\ifx\csname natexlab\endcsname\relax\def\natexlab#1{#1}\fi
\expandafter\ifx\csname bibnamefont\endcsname\relax
  \def\bibnamefont#1{#1}\fi
\expandafter\ifx\csname bibfnamefont\endcsname\relax
  \def\bibfnamefont#1{#1}\fi
\expandafter\ifx\csname citenamefont\endcsname\relax
  \def\citenamefont#1{#1}\fi
\expandafter\ifx\csname url\endcsname\relax
  \def\url#1{\texttt{#1}}\fi
\expandafter\ifx\csname urlprefix\endcsname\relax\def\urlprefix{URL }\fi
\providecommand{\bibinfo}[2]{#2}
\providecommand{\eprint}[2][]{\url{#2}}

\bibitem[{\citenamefont{\.Zyczkowski et~al.}(2001)\citenamefont{\.Zyczkowski,
  Horodecki, Horodecki, and Horodecki}}]{zyczkowski01}
\bibinfo{author}{\bibfnamefont{K.}~\bibnamefont{\.Zyczkowski}},
  \bibinfo{author}{\bibfnamefont{P.}~\bibnamefont{Horodecki}},
  \bibinfo{author}{\bibfnamefont{M.}~\bibnamefont{Horodecki}},
  \bibnamefont{and}
  \bibinfo{author}{\bibfnamefont{R.}~\bibnamefont{Horodecki}},
  \bibinfo{journal}{Phys. Rev. A} \textbf{\bibinfo{volume}{65}},
  \bibinfo{pages}{012101} (\bibinfo{year}{2001}).

\bibitem[{\citenamefont{Yu and Eberly}(2004)}]{yu04}
\bibinfo{author}{\bibfnamefont{T.}~\bibnamefont{Yu}} \bibnamefont{and}
  \bibinfo{author}{\bibfnamefont{J.~H.} \bibnamefont{Eberly}},
  \bibinfo{journal}{Phys. Rev. Lett.} \textbf{\bibinfo{volume}{93}},
  \bibinfo{pages}{140404} (\bibinfo{year}{2004}).

\bibitem[{\citenamefont{Rajagopal and Rendell}(2001)}]{rajagopal}
\bibinfo{author}{\bibfnamefont{A.~K.} \bibnamefont{Rajagopal}}
  \bibnamefont{and} \bibinfo{author}{\bibfnamefont{R.~W.}
  \bibnamefont{Rendell}}, \bibinfo{journal}{Phys. Rev. A}
  \textbf{\bibinfo{volume}{63}}, \bibinfo{pages}{022116}
  (\bibinfo{year}{2001}).

\bibitem[{\citenamefont{Eberly and Yu}(2007)}]{eberly07}
\bibinfo{author}{\bibfnamefont{J.~H.} \bibnamefont{Eberly}} \bibnamefont{and}
  \bibinfo{author}{\bibfnamefont{T.}~\bibnamefont{Yu}},
  \bibinfo{journal}{Science} \textbf{\bibinfo{volume}{316}},
  \bibinfo{pages}{555} (\bibinfo{year}{2007}).

\bibitem[{\citenamefont{\.Zyczkowski et~al.}(1998)\citenamefont{\.Zyczkowski,
  Horodecki, Sanpera, and Lewenstein}}]{zyczkowski98}
\bibinfo{author}{\bibfnamefont{K.}~\bibnamefont{\.Zyczkowski}},
  \bibinfo{author}{\bibfnamefont{P.}~\bibnamefont{Horodecki}},
  \bibinfo{author}{\bibfnamefont{A.}~\bibnamefont{Sanpera}}, \bibnamefont{and}
  \bibinfo{author}{\bibfnamefont{M.}~\bibnamefont{Lewenstein}},
  \bibinfo{journal}{Phys. Rev. A} \textbf{\bibinfo{volume}{58}},
  \bibinfo{pages}{883} (\bibinfo{year}{1998}).

\bibitem[{\citenamefont{Ficek and Tana{\'s}}(2006)}]{ficek06}
\bibinfo{author}{\bibfnamefont{Z.}~\bibnamefont{Ficek}} \bibnamefont{and}
  \bibinfo{author}{\bibfnamefont{R.}~\bibnamefont{Tana{\'s}}},
  \bibinfo{journal}{Phys. Rev. A} \textbf{\bibinfo{volume}{74}},
  \bibinfo{pages}{024304} (\bibinfo{year}{2006}).

\bibitem[{\citenamefont{Sainz and Bj{\"o}rk}(2008)}]{sainz08}
\bibinfo{author}{\bibfnamefont{I.}~\bibnamefont{Sainz}} \bibnamefont{and}
  \bibinfo{author}{\bibfnamefont{G.}~\bibnamefont{Bj{\"o}rk}},
  \bibinfo{journal}{Phys. Rev. A} \textbf{\bibinfo{volume}{77}},
  \bibinfo{pages}{052307} (\bibinfo{year}{2008}).

\bibitem[{\citenamefont{Yu and Eberly}(2009)}]{yu09}
\bibinfo{author}{\bibfnamefont{T.}~\bibnamefont{Yu}} \bibnamefont{and}
  \bibinfo{author}{\bibfnamefont{J.~H.} \bibnamefont{Eberly}},
  \bibinfo{journal}{Science} \textbf{\bibinfo{volume}{323}},
  \bibinfo{pages}{598} (\bibinfo{year}{2009}).

\bibitem[{\citenamefont{Almeida et~al.}(2007)\citenamefont{Almeida, de~Melo,
  Hor-Meyll, Salles, Walborn, Ribeiro, and Davidovich}}]{almeida07}
\bibinfo{author}{\bibfnamefont{M.~P.} \bibnamefont{Almeida}},
  \bibinfo{author}{\bibfnamefont{F.}~\bibnamefont{de~Melo}},
  \bibinfo{author}{\bibfnamefont{M.}~\bibnamefont{Hor-Meyll}},
  \bibinfo{author}{\bibfnamefont{A.}~\bibnamefont{Salles}},
  \bibinfo{author}{\bibfnamefont{S.~P.} \bibnamefont{Walborn}},
  \bibinfo{author}{\bibfnamefont{P.~H.~S.} \bibnamefont{Ribeiro}},
  \bibnamefont{and}
  \bibinfo{author}{\bibfnamefont{L.}~\bibnamefont{Davidovich}},
  \bibinfo{journal}{Science} \textbf{\bibinfo{volume}{316}},
  \bibinfo{pages}{579} (\bibinfo{year}{2007}).

\bibitem[{\citenamefont{Stace and Barrett}(2004)}]{stace04b}
\bibinfo{author}{\bibfnamefont{T.~M.} \bibnamefont{Stace}} \bibnamefont{and}
  \bibinfo{author}{\bibfnamefont{S.~D.} \bibnamefont{Barrett}},
  \bibinfo{journal}{Phys. Rev. Lett.} \textbf{\bibinfo{volume}{92}},
  \bibinfo{pages}{136802} (\bibinfo{year}{2004}).

\bibitem[{\citenamefont{Stace et~al.}(2004)\citenamefont{Stace, Barrett, Goan,
  and Milburn}}]{stace04a}
\bibinfo{author}{\bibfnamefont{T.~M.} \bibnamefont{Stace}},
  \bibinfo{author}{\bibfnamefont{S.~D.} \bibnamefont{Barrett}},
  \bibinfo{author}{\bibfnamefont{H.-S.} \bibnamefont{Goan}}, \bibnamefont{and}
  \bibinfo{author}{\bibfnamefont{G.~J.} \bibnamefont{Milburn}},
  \bibinfo{journal}{Phys. Rev. B} \textbf{\bibinfo{volume}{70}},
  \bibinfo{pages}{205342} (\bibinfo{year}{2004}).

\bibitem[{\citenamefont{Horodecki et~al.}(1999)\citenamefont{Horodecki,
  Horodecki, and Horodecki}}]{horodecki99}
\bibinfo{author}{\bibfnamefont{R.}~\bibnamefont{Horodecki}},
  \bibinfo{author}{\bibfnamefont{M.}~\bibnamefont{Horodecki}},
  \bibnamefont{and}
  \bibinfo{author}{\bibfnamefont{P.}~\bibnamefont{Horodecki}},
  \bibinfo{journal}{Phys. Rev. A} \textbf{\bibinfo{volume}{59}},
  \bibinfo{pages}{1799} (\bibinfo{year}{1999}).

\bibitem[{\citenamefont{Eisert et~al.}(2007)\citenamefont{Eisert, Brandao, and
  Audenaert}}]{eisert}
\bibinfo{author}{\bibfnamefont{J.}~\bibnamefont{Eisert}},
  \bibinfo{author}{\bibfnamefont{F.}~\bibnamefont{Brandao}}, \bibnamefont{and}
  \bibinfo{author}{\bibfnamefont{K.}~\bibnamefont{Audenaert}},
  \bibinfo{journal}{New J. Phys.} \textbf{\bibinfo{volume}{9}},
  \bibinfo{pages}{46} (\bibinfo{year}{2007}).

\bibitem[{\citenamefont{G{\"u}hne et~al.}(2007)\citenamefont{G{\"u}hne,
  Reimpell, and Werner}}]{guhne07a}
\bibinfo{author}{\bibfnamefont{O.}~\bibnamefont{G{\"u}hne}},
  \bibinfo{author}{\bibfnamefont{M.}~\bibnamefont{Reimpell}}, \bibnamefont{and}
  \bibinfo{author}{\bibfnamefont{R.~F.} \bibnamefont{Werner}},
  \bibinfo{journal}{Phys. Rev. Lett.} \textbf{\bibinfo{volume}{98}},
  \bibinfo{pages}{110502} (\bibinfo{year}{2007}).

\bibitem[{\citenamefont{Audenaert and Plenio}(2006)}]{audenaert06}
\bibinfo{author}{\bibfnamefont{K.}~\bibnamefont{Audenaert}} \bibnamefont{and}
  \bibinfo{author}{\bibfnamefont{M.}~\bibnamefont{Plenio}},
  \bibinfo{journal}{New J. Phys.} \textbf{\bibinfo{volume}{8}},
  \bibinfo{pages}{266} (\bibinfo{year}{2006}).

\bibitem[{\citenamefont{G{\"u}hne et~al.}(2008)\citenamefont{G{\"u}hne,
  Reimpell, and Werner}}]{guhne08}
\bibinfo{author}{\bibfnamefont{O.}~\bibnamefont{G{\"u}hne}},
  \bibinfo{author}{\bibfnamefont{M.}~\bibnamefont{Reimpell}}, \bibnamefont{and}
  \bibinfo{author}{\bibfnamefont{R.~F.} \bibnamefont{Werner}},
  \bibinfo{journal}{Phys. Rev. A} \textbf{\bibinfo{volume}{77}},
  \bibinfo{pages}{052317} (\bibinfo{year}{2008}).

\bibitem[{\citenamefont{Puentes et~al.}(2009)\citenamefont{Puentes, Datta,
  Feito, Eisert, Plenio, and Walmsley}}]{puentes09}
\bibinfo{author}{\bibfnamefont{G.}~\bibnamefont{Puentes}},
  \bibinfo{author}{\bibfnamefont{A.}~\bibnamefont{Datta}},
  \bibinfo{author}{\bibfnamefont{A.}~\bibnamefont{Feito}},
  \bibinfo{author}{\bibfnamefont{J.}~\bibnamefont{Eisert}},
  \bibinfo{author}{\bibfnamefont{M.}~\bibnamefont{Plenio}}, \bibnamefont{and}
  \bibinfo{author}{\bibfnamefont{I.}~\bibnamefont{Walmsley}}
  (\bibinfo{year}{2009}), \bibinfo{note}{arXiv:0911.2482 [quant-ph]}.

\bibitem[{\citenamefont{Schmid et~al.}(2008)\citenamefont{Schmid, Kiesel,
  Wieczorek, and Weinfurter}}]{schmid08}
\bibinfo{author}{\bibfnamefont{C.}~\bibnamefont{Schmid}},
  \bibinfo{author}{\bibfnamefont{N.}~\bibnamefont{Kiesel}},
  \bibinfo{author}{\bibfnamefont{W.}~\bibnamefont{Wieczorek}},
  \bibnamefont{and}
  \bibinfo{author}{\bibfnamefont{H.}~\bibnamefont{Weinfurter}},
  \bibinfo{journal}{Phys. Rev. Lett.} \textbf{\bibinfo{volume}{101}},
  \bibinfo{pages}{260505} (\bibinfo{year}{2008}).

\bibitem[{\citenamefont{Vagov et~al.}(2003)\citenamefont{Vagov, Axt, and
  Kuhn}}]{vagov03}
\bibinfo{author}{\bibfnamefont{A.}~\bibnamefont{Vagov}},
  \bibinfo{author}{\bibfnamefont{V.~M.} \bibnamefont{Axt}}, \bibnamefont{and}
  \bibinfo{author}{\bibfnamefont{T.}~\bibnamefont{Kuhn}},
  \bibinfo{journal}{Phys. Rev. B} \textbf{\bibinfo{volume}{67}},
  \bibinfo{pages}{115338} (\bibinfo{year}{2003}).

\bibitem[{\citenamefont{Vagov et~al.}(2004)\citenamefont{Vagov, Axt, Kuhn,
  Langbein, Borri, and Woggon}}]{vagov04}
\bibinfo{author}{\bibfnamefont{A.}~\bibnamefont{Vagov}},
  \bibinfo{author}{\bibfnamefont{V.~M.} \bibnamefont{Axt}},
  \bibinfo{author}{\bibfnamefont{T.}~\bibnamefont{Kuhn}},
  \bibinfo{author}{\bibfnamefont{W.}~\bibnamefont{Langbein}},
  \bibinfo{author}{\bibfnamefont{P.}~\bibnamefont{Borri}}, \bibnamefont{and}
  \bibinfo{author}{\bibfnamefont{U.}~\bibnamefont{Woggon}},
  \bibinfo{journal}{Phys. Rev. B} \textbf{\bibinfo{volume}{70}},
  \bibinfo{pages}{201305(R)} (\bibinfo{year}{2004}).

\bibitem[{\citenamefont{Borri et~al.}(2001)\citenamefont{Borri, Langbein,
  Schneider, Woggon, Sellin, Ouyang, and Bimberg}}]{borri01}
\bibinfo{author}{\bibfnamefont{P.}~\bibnamefont{Borri}},
  \bibinfo{author}{\bibfnamefont{W.}~\bibnamefont{Langbein}},
  \bibinfo{author}{\bibfnamefont{S.}~\bibnamefont{Schneider}},
  \bibinfo{author}{\bibfnamefont{U.}~\bibnamefont{Woggon}},
  \bibinfo{author}{\bibfnamefont{R.~L.} \bibnamefont{Sellin}},
  \bibinfo{author}{\bibfnamefont{D.}~\bibnamefont{Ouyang}}, \bibnamefont{and}
  \bibinfo{author}{\bibfnamefont{D.}~\bibnamefont{Bimberg}},
  \bibinfo{journal}{Phys. Rev. Lett.} \textbf{\bibinfo{volume}{87}},
  \bibinfo{pages}{157401} (\bibinfo{year}{2001}).

\bibitem[{\citenamefont{Roszak and
  Machnikowski}(2006{\natexlab{a}})}]{roszak06a}
\bibinfo{author}{\bibfnamefont{K.}~\bibnamefont{Roszak}} \bibnamefont{and}
  \bibinfo{author}{\bibfnamefont{P.}~\bibnamefont{Machnikowski}},
  \bibinfo{journal}{Phys. Rev. A} \textbf{\bibinfo{volume}{73}},
  \bibinfo{pages}{022313} (\bibinfo{year}{2006}{\natexlab{a}}).

\bibitem[{\citenamefont{Roszak and Machnikowski}(2009)}]{roszak09}
\bibinfo{author}{\bibfnamefont{K.}~\bibnamefont{Roszak}} \bibnamefont{and}
  \bibinfo{author}{\bibfnamefont{P.}~\bibnamefont{Machnikowski}},
  \bibinfo{journal}{Phys. Rev. B} \textbf{\bibinfo{volume}{80}},
  \bibinfo{pages}{195315} (\bibinfo{year}{2009}).

\bibitem[{\citenamefont{Roszak and
  Machnikowski}(2006{\natexlab{b}})}]{roszak05a}
\bibinfo{author}{\bibfnamefont{K.}~\bibnamefont{Roszak}} \bibnamefont{and}
  \bibinfo{author}{\bibfnamefont{P.}~\bibnamefont{Machnikowski}},
  \bibinfo{journal}{Phys. Lett. A} \textbf{\bibinfo{volume}{351}},
  \bibinfo{pages}{251} (\bibinfo{year}{2006}{\natexlab{b}}).

\bibitem[{\citenamefont{Krummheuer et~al.}(2002)\citenamefont{Krummheuer, Axt,
  and Kuhn}}]{krummheuer02}
\bibinfo{author}{\bibfnamefont{B.}~\bibnamefont{Krummheuer}},
  \bibinfo{author}{\bibfnamefont{V.~M.} \bibnamefont{Axt}}, \bibnamefont{and}
  \bibinfo{author}{\bibfnamefont{T.}~\bibnamefont{Kuhn}},
  \bibinfo{journal}{Phys. Rev. B} \textbf{\bibinfo{volume}{65}},
  \bibinfo{pages}{195313} (\bibinfo{year}{2002}).

\bibitem[{\citenamefont{Hill and Wootters}(1997)}]{hill97}
\bibinfo{author}{\bibfnamefont{S.}~\bibnamefont{Hill}} \bibnamefont{and}
  \bibinfo{author}{\bibfnamefont{W.~K.} \bibnamefont{Wootters}},
  \bibinfo{journal}{Phys. Rev. Lett.} \textbf{\bibinfo{volume}{78}},
  \bibinfo{pages}{5022} (\bibinfo{year}{1997}).

\bibitem[{\citenamefont{Wootters}(1998)}]{wootters98}
\bibinfo{author}{\bibfnamefont{W.~K.} \bibnamefont{Wootters}},
  \bibinfo{journal}{Phys. Rev. Lett.} \textbf{\bibinfo{volume}{80}},
  \bibinfo{pages}{2245} (\bibinfo{year}{1998}).

\bibitem[{\citenamefont{Chuang and Nielsen}(1997)}]{chuang97}
\bibinfo{author}{\bibfnamefont{I.~L.} \bibnamefont{Chuang}} \bibnamefont{and}
  \bibinfo{author}{\bibfnamefont{M.~A.} \bibnamefont{Nielsen}},
  \bibinfo{journal}{J. of Mod. Opt.} \textbf{\bibinfo{volume}{44}},
  \bibinfo{pages}{2455} (\bibinfo{year}{1997}).

\bibitem[{\citenamefont{Riebe et~al.}(2006)\citenamefont{Riebe, Kim, Schindler,
  Monz, Schmidt, K{\"o}rber, H{\"a}nsel, H{\"a}ffner, Roos, and
  Blatt}}]{riebe06}
\bibinfo{author}{\bibfnamefont{M.}~\bibnamefont{Riebe}},
  \bibinfo{author}{\bibfnamefont{K.}~\bibnamefont{Kim}},
  \bibinfo{author}{\bibfnamefont{P.}~\bibnamefont{Schindler}},
  \bibinfo{author}{\bibfnamefont{T.}~\bibnamefont{Monz}},
  \bibinfo{author}{\bibfnamefont{P.~O.} \bibnamefont{Schmidt}},
  \bibinfo{author}{\bibfnamefont{T.~K.} \bibnamefont{K{\"o}rber}},
  \bibinfo{author}{\bibfnamefont{W.}~\bibnamefont{H{\"a}nsel}},
  \bibinfo{author}{\bibfnamefont{H.}~\bibnamefont{H{\"a}ffner}},
  \bibinfo{author}{\bibfnamefont{C.~F.} \bibnamefont{Roos}}, \bibnamefont{and}
  \bibinfo{author}{\bibfnamefont{R.}~\bibnamefont{Blatt}},
  \bibinfo{journal}{Phys. Rev. Lett.} \textbf{\bibinfo{volume}{97}},
  \bibinfo{pages}{220407} (\bibinfo{year}{2006}).

\bibitem[{\citenamefont{Zanardi et~al.}(2000)\citenamefont{Zanardi, Zalka, and
  Faoro}}]{zanardi00}
\bibinfo{author}{\bibfnamefont{P.}~\bibnamefont{Zanardi}},
  \bibinfo{author}{\bibfnamefont{C.}~\bibnamefont{Zalka}}, \bibnamefont{and}
  \bibinfo{author}{\bibfnamefont{L.}~\bibnamefont{Faoro}},
  \bibinfo{journal}{Phys. Rev. A} \textbf{\bibinfo{volume}{62}},
  \bibinfo{pages}{030301(R)} (\bibinfo{year}{2000}).

\bibitem[{\citenamefont{Jamio{\l}kowski}(1972)}]{jamiolkowski72}
\bibinfo{author}{\bibfnamefont{A.}~\bibnamefont{Jamio{\l}kowski}},
  \bibinfo{journal}{Rep. Math. Phys.} \textbf{\bibinfo{volume}{3}},
  \bibinfo{pages}{275} (\bibinfo{year}{1972}).

\bibitem[{\citenamefont{Choi}(1982)}]{choi82}
\bibinfo{author}{\bibfnamefont{M.-D.} \bibnamefont{Choi}},
  \bibinfo{journal}{Proc. Symp. Pure Math.} \textbf{\bibinfo{volume}{38}},
  \bibinfo{pages}{583} (\bibinfo{year}{1982}).

\end{thebibliography}
\end{document}